\documentclass[fleqn,twoside,twocolumn,nofootinbib]{revtex4} 
\usepackage{ujp} 
\begin{document}
\title[THICKNESS DEPENDENCES OF PHOTOELECTRIC CHARACTERISTICS]
{THICKNESS DEPENDENCES OF PHOTOELECTRIC CHARACTERISTICS OF SILICON
BACKSIDE\\ CONTACT SOLAR CELLS
}%
\author{А.P.~Gorban }
\affiliation{V. Lashkaryov Institute of Semiconductor Physics, Nat. Acad. of Sci. of Ukraine}
\address{41, Prosp. Nauky, Kyiv 03680, Ukraine}
\email{sach@isp.kiev.ua}
\author{V.P.~Kostylyov}
\affiliation{V. Lashkaryov Institute of Semiconductor Physics, Nat. Acad. of Sci. of Ukraine}
\address{41, Prosp. Nauky, Kyiv 03680, Ukraine}
\email{sach@isp.kiev.ua}
\author{А.V.~Sachenko }
\affiliation{V. Lashkaryov Institute of Semiconductor Physics, Nat. Acad. of Sci. of Ukraine}
\address{41, Prosp. Nauky, Kyiv 03680, Ukraine}
\email{sach@isp.kiev.ua}
\author{О.А.~Serba }
\affiliation{V. Lashkaryov Institute of Semiconductor Physics, Nat. Acad. of Sci. of Ukraine}
\address{41, Prosp. Nauky, Kyiv 03680, Ukraine}
\email{sach@isp.kiev.ua}
\author{І.О.~Sokolovskyi}
\affiliation{V. Lashkaryov Institute of Semiconductor Physics, Nat. Acad. of Sci. of Ukraine}
\address{41, Prosp. Nauky, Kyiv 03680, Ukraine}
\email{sach@isp.kiev.ua}
\author{V.V.~Chernenko}
\affiliation{V. Lashkaryov Institute of Semiconductor Physics, Nat. Acad. of Sci. of Ukraine}
\address{41, Prosp. Nauky, Kyiv 03680, Ukraine}
\email{sach@isp.kiev.ua}

\udk{621.315.592} \pacs{71.55.Cn, 72.20.Jv,\\[-3pt] 72.40.+w} \razd{\secix}

\setcounter{page}{161}%
\maketitle

\begin{abstract}
The thickness dependences of the photocurrent quantum yield and
photoenergy parameters of silicon backside contact solar cells (BC
SC) are investigated theoretically and experimentally. The surface
recombination rate on the irradiated surface was minimized by means
of creating the layers of microporous silicon. A method of finding
the surface recombination rate and the diffusion length of minority
carriers from the thickness dependences of the photocurrent quantum
yield under conditions of the strong absorption is proposed. The
performed studies allowed us to establish that the thinning of the
BC SC samples in the case of minimizing the surface recombination
rate gives a possibility to achieve rather high efficiencies of
photoconversion. It is also shown that the agreement between the
experimental and theoretical spectral dependences of the
photocurrent quantum yield can be reached only with regard for the
coefficient of light reflection from the backside surface.
\end{abstract}

\section{Introduction}

Silicon backside contact  (BC) solar cells (SC) with $n$-type base,
as well as solar batteries produced on their basis, have the highest
efficiency of photoelectrical energy conversion $\eta$ achieved for
today that reaches 20 \% for serial modules \cite{1}. The thickness
of the quasineutral base region in such BC SCs is usually much
smaller than the diffusion length of minority carriers, whereas the
effective surface recombination rate $S^*$ on the front
(nonmetallized) surface referred to the inner boundary of the
near-surface space charge region (SCR) is minimized to the level
having practically no influence on the value of $\eta$.

As is known, the most effective way of eliminating the surface
recombination losses in BC SCs is to generate isotype $n^+-n$ or
$p^+-p$ junctions on their front surface that limit the supply of
nonequilibrium minority carriers to surface recombination centers
\cite{2,3}. In the presence of such junctions, the effective surface
recombination rate $S^*$ is minimized due to a decrease of the
minority carrier fluxes via surface recombination centers \cite{2}.
Though, at high doping levels of the surface layer, the rate
$S^*$ can increase due to an increase of the Auger recombination rate in
it \cite{3}.

A number of works \cite{4,5} used another way  of minimizing the
negative influence of surface recombination losses on the BC SC
efficiency $\eta$, namely the formation of floating $p^+-n$ or
$n^+-p$ junctions on their front surface that limited the supply of
nonequilibrium majority carriers to surface recombination centers.
However, the experimental researches \cite{5} demonstrated that,
though the formation of a floating $n^+-p$ junction really resulted
in an increase of the BC SC efficiency $\eta$ under the standard
spectral АМ1.5 conditions at the irradiance $P =1000$ W/m$^2$, a
considerable (tens-fold) rise of the effective surface recombination
rate $S^*$ was simultaneously observed. As was shown in our work
\cite{6}, the increase of $S^*$ is related to the contribution of
the SCR recombination that can be very significant in the case of
low-intensity irradiance.\looseness=1

If the initial thickness of a BC SC exceeds the diffusion length of
minority carriers, then its thinning must result in a considerable
rise of the short-circuit current and the photoconversion
efficiency. This work is devoted to experimental and theoretical
studies of the thickness dependences of the quantum yield and
photoenergy parameters of BC SCs. The effective surface
recombination rate after a regular thinning of a BC SC sample was
minimized by means of creating a microporous silicon layer on the
sample surface. We propose a method of determination of the
effective surface recombination rate $S^*$ and the diffusion length
of minority carriers $L$ from the thickness dependences of the BC SC
quantum yield under conditions of the strong light absorption in a
semiconductor. It is established that the long-wavelength maximum at
the spectral dependences of the BC SC short-circuit current
perceptibly depends on the coefficient of light reflection from the
BC SC backside metallized surface.\looseness=1

\begin{figure}
\includegraphics[width=8cm]{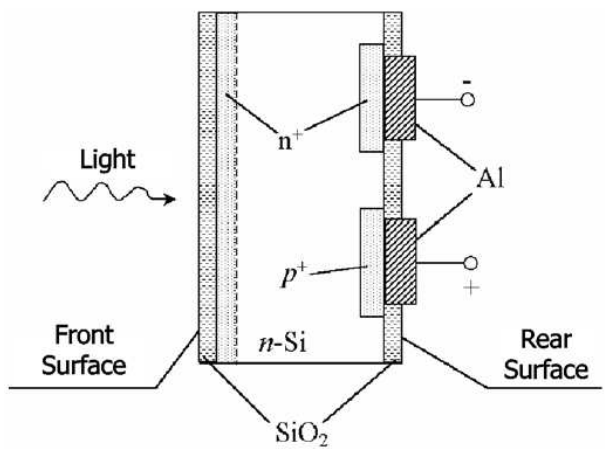}
\vskip-3mm\caption{Diagram of a silicon backside contact solar cell}
\end{figure}

\section{Experimental Technique}

The recombination activity was studied using experimental samples of
silicon BC SCs from two groups with initial thicknesses of 400
$\mu$m. Their schematic cross section is shown in Fig.~1. The
samples were produced on plates of $n$-type zone-melting silicon
with the resistivity $\rho=2$ Ohm$\cdot$cm. A near-surface isotype
$n^+-n$ junction or a floating $p^+-n$ junction was formed on the
front (irradiated) surface 2 cm$^2$ in area. The front surface of
the BC SCs was additionally passivated by a thermal SiO$_2$ layer
with a thickness approximating 110 nm, which reduced the optical losses of
incident light and the concentration of surface recombination-active
centers on this surface. The BC SC samples without near-surface
junctions with the only thermal SiO$_2$ layer on the front surface
were also investigated.

The choice of BC SCs as an instrument for studying the nature of
surface and volume recombination processes is caused by the fact
that they allow one to rather easily realize the conditions, under
which the region of optical generation of nonequilibrium
electron-hole pairs will be localized close to the front surface and
spatially separated from the collector junction by the quasineutral
base region. The short-circuit current of the collector junction on
the backside surface is a function of the effective surface
recombination rate $S^*$ on the front surface that depends, in turn,
on the rates of recombination via surface recombination-active
centers, Auger recombination in the heavily doped $n^+$- or
$p^+$-layers, and recombination in the near-surface SCR.
Investigating the kinetics of variation of the short-circuit current
of the collector junction in the process of the BC SC thinning, it
is possible to separate the contributions made to the photocurrent
by the surface and volume recombinations.

The experimental BC SC samples with Al buses  welded to the
contact areas were firstly thermally fixed on the surface of a
glass substrate with the help of an optically transparent butyral
resin film. After that, the Al buses were welded with transfer
metal electrodes subsequently soldered with conducting metal
wires. All the components of the set-up were protected from the
action of aggressive chemical substances of an electrolyte or
etchant by chemical-resistant optically transparent polymer
materials, so that the electrolyte or etchant could contact only
with the front thermally oxidized BC SC surface.

The chemical etching was performed in a mixture of HNO$_3$ and HF
relating as 3:1. At the chemical etching, a silicon dioxide layer
close to the BC SC front surface was etched first of all, then
followed the isotype $n^+-n$ or the floating $p^+-n$ junction, and
only after that took place the etching of the base region. The
thickness of the etched layers was determined by the etching time
and controlled by a micrometer.

The processes of anode etching of the BC SC front  surface were
performed in a transparent electrochemical cell with a Pt electrode.
The cell construction allowed one to irradiate the BC SC front
surface by light coming from a mirror incandescent lamp through a
selective optical filter SZS-26 that transmitted only photons with
$\lambda<0.75$ $\mu$m and thus provided a small (several micrometers)
effective depth of optical generation of electron-hole pairs close
to the BC SC front surface. The irradiance of the BC SC surface in
the cell in the presence of an electrolyte approximated 700 W/m$^2$.

The electrochemical anodic process was performed in the
galvanostatic mode at a constant current density on the sample
surface ($J = $~2--4 mA/cm$^2$) and the voltage at the output of a
power supply $V = 2$ V. The used electrolyte was a mixture of
ethanol and concentrated (49\%) HF taken in the ratio of 1:4.
Besides the stabilized current density and the voltage across the
electrolyte cell, we also controlled the short-circuit current of
the backside collector junction caused by the irradiation of the BC
SC front surface. The latter allowed us to determine the dynamics of
the variation of recombination parameters of the BC SC front surface
directly in the course of the electrochemical reaction.

Each thinning of the BC SC sample was followed by the anodization
leading to the formation of a microporous silicon layer on the
surface. This resulted in both the passivation of the surface and a decrease
of the surface recombination rate. After etching the silicon
dioxide and the heavily doped layers, the effective surface
recombination rate $S^*$ grew due to the formation of
depleting band bendings on the front surface, whereas the
anodization of the etched surface led to a decrease of $S^*$ due to
the hydrogen passivation of the surface.

The phototechnical and optical parameters of the BC SCs  were
measured with the help of a control-measuring equipment of the
Center for testing photoconverters and photoelectric batteries of
V.E. Lashkaryov Institute of Semiconductor Physics of the NAS of
Ukraine certified by the State Committee of Ukraine for Technical
Regulation and Consumer Policy. The measuring technique was as
follows. First, we determined the initial phototechnical parameters
of the BC SCs with a thermally oxidized surface under the AM1.5
spectral conditions and measured the spectral dependences of the
short-circuit current in the wavelength range $\Delta\lambda =
0.4...1.2$ $\mu$m. Then, after each thinning and the following
anodization, the BC SC sample was dried at room temperature, and the
same parameters were measured once again.

\section{Thickness Dependences of the Internal Quantum Yield of BC SCs}

The internal quantum yield $Q$ in silicon BC SCs in the case of
monochromatic irradiation can be found from the solution of a
diffusion equation with the following boundary conditions:
\begin{equation}
J(x=0)=-S^*\Delta p(x=0),
\end{equation}
\begin{equation}
\Delta p(x=d)=0,
\end{equation}
\noindent where $j(x)$ and $\Delta p(x)$ are the flux of
electron-hole pairs and their excess concentration in the plane $x$,
respectively, $S^*$ is the total rate of surface recombination $S$
and recombination in the SCR $V_{\rm SC}$ in the plane $x=w$, where
$w$ is the SCR thickness.

Using the solution of the standard diffusion equation for excess
electron-hole pairs with regard for the light reflection from the
backside surface, we obtain the following expression for the
internal quantum yield $Q$:
\[
Q=\frac{\alpha L}{1-\alpha^2 L^2}\times
\]
\[
 \times
\Bigl\{-\Bigl[\frac{S^* L}{D} \left((1+R_d e^{-2 \alpha
d})-(1+R_d)e^{-\alpha d-d/L}\right)+
\]
\[
+\alpha L (1-R_d e^{-2\alpha d})+(1+R_d)e^{-\alpha
d-d/L}\Bigr]\times
\]
\[
\times\left[{\rm cosh}(\frac{d}{L})+\frac{S^* L}{D} {\rm
sinh}(\frac{d}{L})\right]^{-1}+
\]
\begin{equation}
 +\left[ (1+R_d)+\alpha L (1-R_d)
\right] e^{-\alpha d} \Bigr\},
\end{equation}
\noindent where $L=\sqrt{D\tau}$ stands for the diffusion length of
minority carriers, $D$ is the diffusion coefficient, $\alpha$ is the
light absorption coefficient, and $R_d$ is the coefficient of light
reflection from the backside surface.

The authors of \cite{7} also obtained an expression for the
spectral dependence of the BC SC internal quantum yield with the use
of boundary conditions (1) and (2) and proposed a method of deriving
the surface recombination rate $S^*$ and the diffusion length of
minority carriers $L$ based on these spectral dependences. However,
that study was performed without regard for the effect of light
reflection from the backside surface.

In the case of strong light absorption ($\alpha L\gg 1$ and $\alpha
d\gg 1$), the expression for the internal quantum yield of BC SCs
significantly simplifies and takes the form
\begin{equation}
Q=\left(\cos h\left(\frac{d}{L}\right)+\frac{S^* L}{D}\sin
h\left(\frac{d}{L}\right)\right)^{-1}.
\end{equation}
As one can see from (4), the internal quantum yield in the case of
strong light absorption does not depend on the coefficients of light
absorption $\alpha$ and light reflection from the backside surface
$R_d$.

In the case where the inequality $d\ll L$ is satisfied, Eq.(4) yields
\begin{equation}
Q=\left(1+\frac{S^* d}{D}\right)^{-1}.
\end{equation}
Thus, the inverse quantum yield is equal to
\begin{equation}
\frac{1}{Q}=1+\frac{S^* d}{D},
\end{equation}
\noindent i.e., it represents a straight-line dependence on $d$,
whose slope $S^*/D$ allows one to determine the surface
recombination rate. However, the diffusion length cannot be
found in this case.

\begin{figure}
\includegraphics[width=\column]{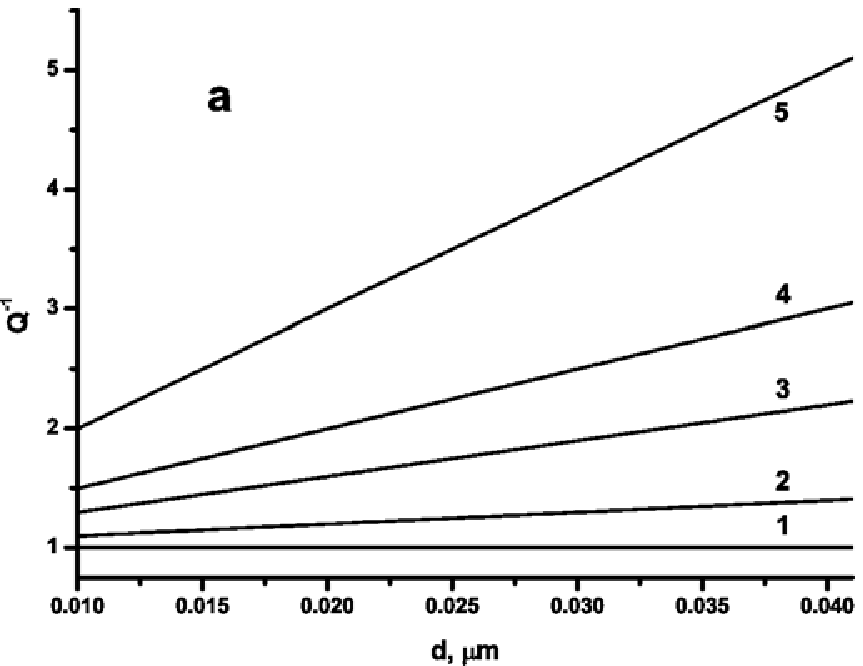}\\[2mm]
\includegraphics[width=\column]{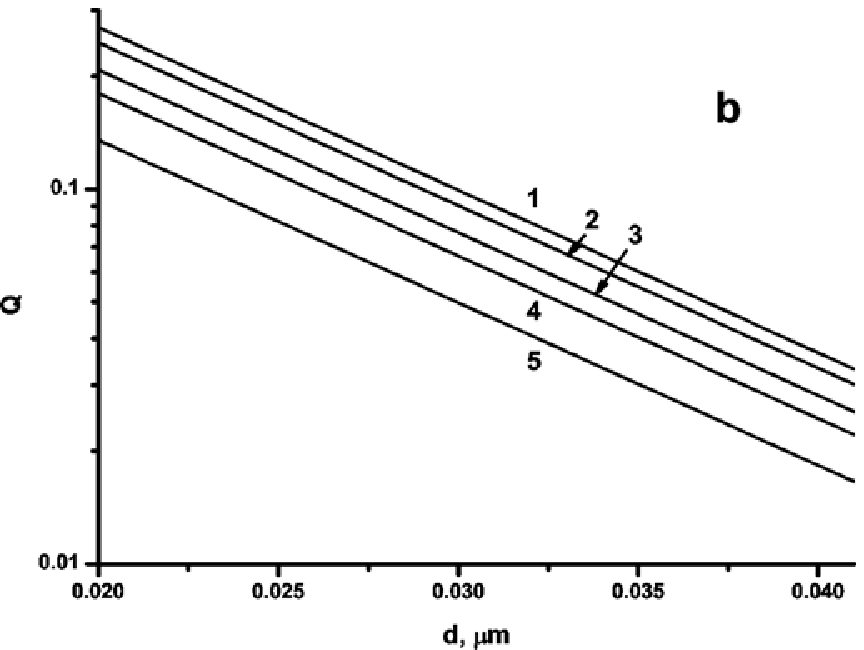}
\vskip-3mm\caption{ Theoretical dependences of the inverse quantum yield
 (\textit{a}) and quantum yield ({\it b}) on the thickness in the case of strong light absorption for the limiting
  cases of thick and thin BS SCs as compared to the diffusion length at $L=100$~$\mu$m
  ({\it 1} ({\it b})) and $S^*$ = 1 cm/s (\textit{1}), $10^2$ cm/s (\textit{2}), $3 \times 10^2$ cm/s (\textit{3}),
  $5 \times 10^2$ (\textit{4}), and $10^3$ (\textit{5})}
\end{figure}

The corresponding dependences are presented in Fig.~2,{\it a}.

If $d\gg L$, expression (4) yields
\begin{equation}
Q=\exp\left(-\frac{d}{L}\right)/\left(1+\frac{S^* L}{D}\right).
\end{equation}
\noindent In this case, the thickness dependences of ln$(Q)$ are
linear (Fig.~2,\textit{b}). Moreover,  the thickness dependences of
$Q$ enable one to determine the diffusion length $L$, whereas the
accuracy of deriving the surface recombination rate $S^*$
abruptly decreases, and it can be found only at $S^* L/D\geq 1$. The
simultaneous determination of $L$ and $S^\ast$ from the thickness
dependences of the BC SC internal quantum yield is possible only if
$L \sim d$ and $S^* L/D\geq 1$.

According to (4), if the short-circuit current is measured under
strong light absorption, then the ratio of the short-circuit
currents at the minimum thickness $d_{\min}$ and the arbitrary
thickness $d$ is determined by the expression
\begin{equation}
N(d)=\frac{\cos h(d/L)+S^* L/D \sin h(d/L)} {\cos h(d_{\min}/L)+S^*
L/D \sin h(d_{\min}/L)}.
\end{equation}
The ratio $N$ represents an increasing function of the thickness if
it changes from the minimum to the maximum value. At the known
thickness $d$ and diffusion coefficient $D$, dependence (8) contains
two unknown parameters: $S^*$ and $L$. Unfortunately, in the real
case of $d_{\max}<L$, the dependence $N(d)$ represents an ambiguous
function of $S^*$ and $L$. That is why their definite determination
requires some additional information, for example, obtained from
measurements of spectral dependences of the short-circuit current.

\begin{figure}
\includegraphics[width=8.4cm]{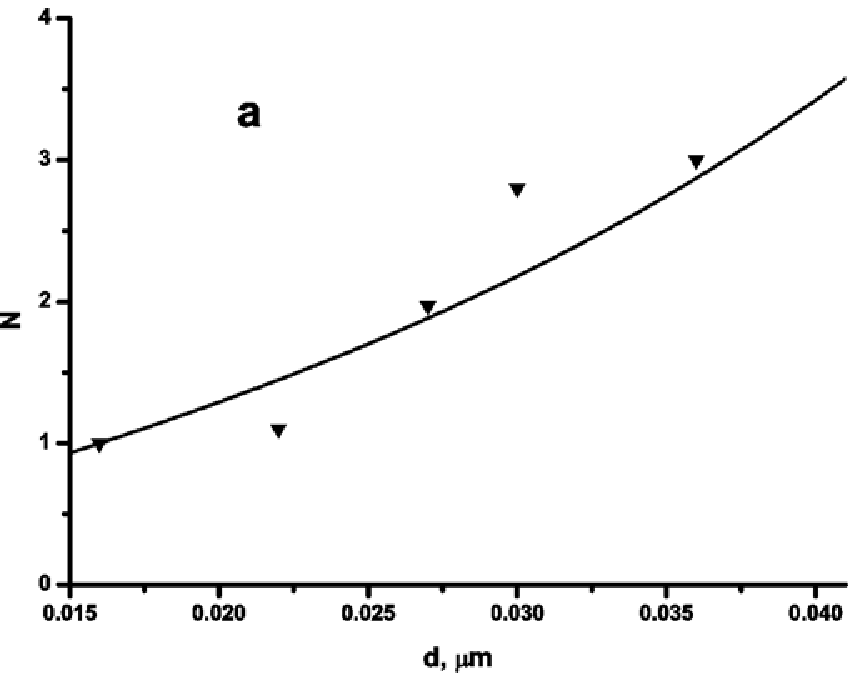}\\[2mm]
\includegraphics[width=8.4cm]{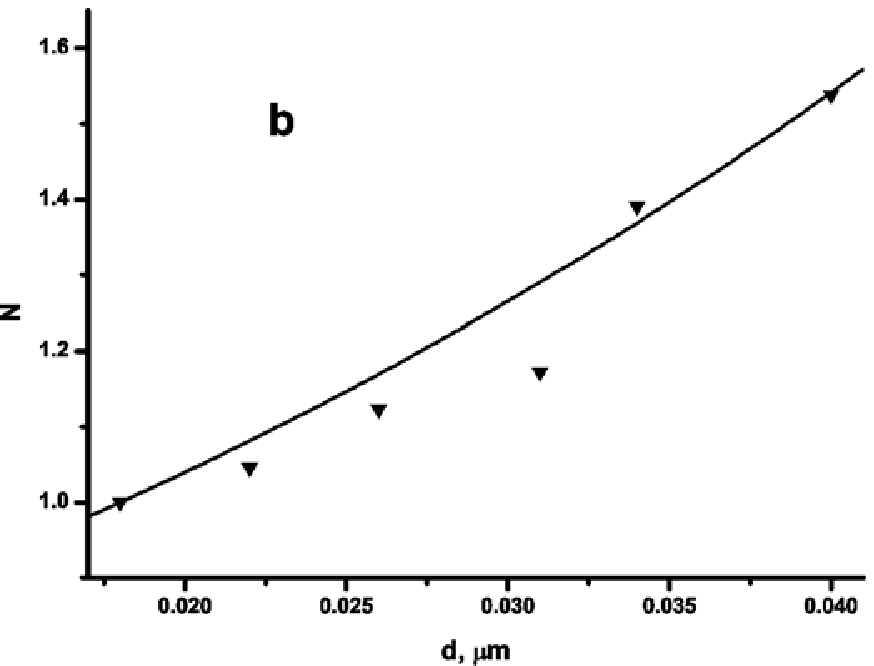}
\vskip-3mm\caption{Experimental (dots) and theoretical (solid lines)
thickness dependences of the normalized quantum yield for the
samples from the first (\textit{а}) and second (\textit{b}) groups.
$L$=250 $\mu$m, $S^*$= $2 \times 10^4$ cm/s (\textit{a}); $L=500$
$\mu$m, $S^*$= 190 cm/s (\textit{b})}
\end{figure}

Figure 3 presents the experimental dependences $N(d)$ obtained for
the BC SC samples from two groups. This figure also shows the
theoretical functions $N(d)$ obtained with the use of expression (8)
and such values of $S^*$ and $L$, at which the experimental
dependences agree with the theoretical ones. The agreement between
the experimental and theoretical dependences is satisfactory.

The experimental spectral dependences of the internal quantum yield
for the BC SC samples from the first group with a floating $p-n$
junction near the front surface are given in Fig.~4. The theoretical
functions are constructed with the use of expression (3) and the
spectral dependence of the absorption coefficient in silicon
$\alpha(\lambda)$ taken from work \cite{8}, where this dependence was
determined at various temperatures. The spectral dependence
$\alpha(\lambda)$ obtained in \cite{8} at 300 K can be numerically
approximated by the expression
\[
\log(\alpha(\lambda))=306.91-25.649\lambda^{-1}-1399.55\lambda+
\]
\[
+3403.54\lambda^2-4782.3\lambda^3+3901.23\lambda^4-
\]
\begin{equation}
-1713.98\lambda^5+311.596\lambda^6\;,
\end{equation}
where the wavelength $\lambda$ is measured in micrometers.

As one can see from Fig.~4, the use of the value  of 250 $\mu$m
for the diffusion length allows one to match the experimental
dependences with the theoretical ones, if the
coefficient of light reflection from the backside
surface is significant ($\sim 0.7$), and the value of $S^*$ which is determined in this case by the
SCR recombination rate is large, which is true at low irradiation levels
realized with the use of a monochromator. Without regard for the
light reflection from the backside surface, the calculated maximum
lies to the left from the experimental one.

The inset in Fig. 4 shows the theoretical spectral dependences of
the short-circuit current at various diffusion lengths $L$. As one
can see from the inset, an increase of the diffusion length
practically does not affect the position of the maximum of the
function $J_{\rm SC}(\lambda)$ but results in a rise of its value.

\begin{figure}
\includegraphics[width=\column]{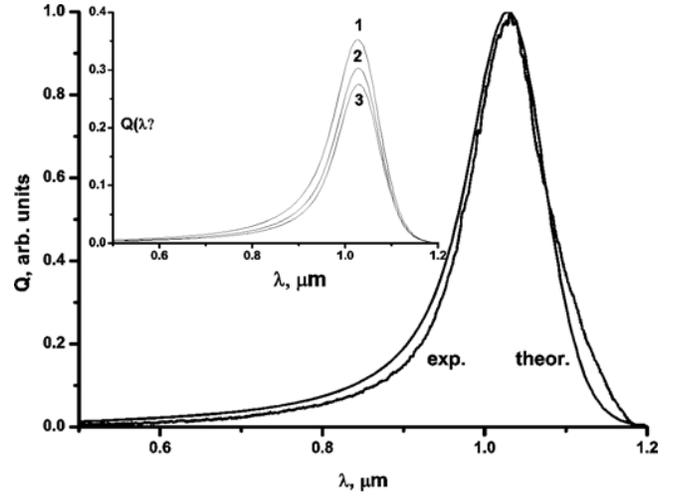}
\vskip-3mm\caption{Experimental and theoretical spectral dependences
of the BC SC quantum yield normalized to its maximum value at
$R_d=0.7$, $S^*=7 \times 10^4$ cm/s, $L=250$ $\mu$m; in the inset --
spectral dependences of the BC SC quantum yield at $R_d=0.7$, $S^*=7
\times 10^4$ cm/s and various $L$: 500 $\mu$m (\textit{1}), 250
$\mu$m (\textit{2}), and 500 $\mu$m (\textit{3}).}
\end{figure}

\section{Thickness Dependences of BC SC Photoenergy Parameters}

Theoretical thickness dependences of such photoenergy parameters as
the  short-circuit current $I_{\rm SC}$, the open-circuit voltage
$V_{\rm OC}$, and the photoconversion efficiency $\eta$ of BC SCs
were simulated, in particular, in \cite{9,10,11}. The photoconversion
efficiency of BC SCs depends most considerably on the thickness
dependence of the short-circuit current. In \cite{11}, it was shown
that, in the case where $d<L$, the thickness dependence of the
short-circuit current density of BC SCs $J_{\rm SC}$ under the
condition of total light trapping is determined by the formula
\begin{equation}
J_{\rm SC}\cong \frac{J_{\rm gen}}{1+S^* d/D_a},
\end{equation}
\noindent where $J_{\rm SC}$ stands for the generation current
density in a silicon BC SC  for the indicated irradiation conditions,
and $D_a$ is the bipolar diffusion coefficient.

In the case of the linearity with respect to the excess
concentration of  electron-hole pairs (i.e. $\Delta p(x=w)<n_0$,
where $n_0$ is the equilibrium concentration of majority carriers in
the BC SC base), one can calculate the short-circuit current density
in BC SCs with the use of expression (3), by taking the light
reflection from the front surface and its incomplete absorption in
the semiconductor into account. In particular, under the АМ0 conditions, for
which the solar radiation spectrum can be approximated to a good
accuracy by the blackbody radiation with a temperature of 5800 K,
the short-circuit current density of a silicon BC SC at room
temperature is described by the expression
\begin{equation}
J_{\rm SC}\cong 0.656 \int\limits^{1,13}_{0}
\frac{(1-R_s(\lambda))Q(\lambda)}{\lambda^4
 (\exp(2,494/\lambda)-1)}d\lambda ,
\end{equation}
\noindent where $R_s(\lambda)$ is the coefficient of light
reflection from the front surface, while the irradiation wavelength
$\lambda$ and the current density are measured in micrometers and
А/cm$^2$, respectively.

\begin{figure}
\includegraphics[width=\column]{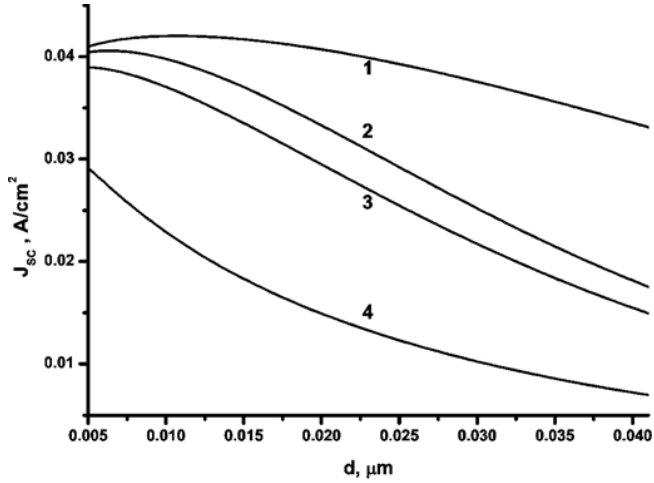}
\vskip-3mm\caption{Calculated thickness dependences of the BC SC
short-circuit  current under the АМ0 conditions. $R_d=0.7$; $L=500$
$\mu$m (\textit{1}), 250 $\mu$m (\textit{2—-4}); $S^*=10$ cm/s
(1,2), $10^2$ (\textit{3}), and $10^3$ (\textit{4})}
\end{figure}

Figure 5 demonstrates the theoretical thickness dependences of the
short-circuit  current density in BC SCs obtained with the use of
expression (11) in the case where $R_s(\lambda)=0.1$. As one can
see from Fig.~5, these dependences have a maximum at sufficiently
small values of $S^*$ (see curves {\it 1, 2}). Moreover, its
position shifts toward lower thicknesses $d$ with increase in the
value of $S^*$ and with decrease in the diffusion length $L$. In the given
case, the existence of the maximum is due to the competition of two
processes. As the thickness decreases, the total recombination in the
quasineutral volume falls, due to which the value of $J_{\rm SC}$
must increase. At the same time, the number of electron-hole pairs
generated by Sun's light also decreases because of a smaller absorption,
which favors a reduction of $J_{\rm SC}$. Depending on which of the
indicated factors dominates, we obtain either an increase or a
decrease of the short-circuit current with decrease in the thickness.
However, in the case of large $S^*$, where the losses due to the surface
recombination play a dominant role, the region, where
$J_{\rm SC}$ decreases with increase in $d,$ is absent (curve {\it 4}). In this
case, the short-circuit current grows with decrease in the thickness even
at very small thicknesses ($\sim 10$ $\mu$m). It is explained by the
fact that, with decrease in the thickness at $d<L$, the majority of
generated pairs will move to the backside surface characterized by a
very high ``recombination rate'' in the short-circuit mode.
Therefore, the value of $J_{\rm SC}$ will rise.

In the case of total light trapping, where $J_{\rm SC}$ satisfies
expression (10), no reduction of the short-circuit current at
small $d$ is present, whereas its value at $d \approx L$ exceeds that
obtained from Eq.~(11).

As was noted above, when searching for the thickness dependence of
the BC SC photoconversion  efficiency $\eta$, the latter is mainly
determined by the function $J_{\rm SC}(d)$. Therefore,
for a unit-area SC under the АМ0 conditions, we can write
\begin{equation}
\eta_{\rm AM0}\approx J_{\rm SC}(d) V_{\rm OC} FF/0.135\;,
\end{equation}
\noindent where $FF$ is the occupation factor of the current-voltage
characteristic, and the factor of $0.135$ W/cm$^2$ is the irradiation power under
the АМ0 conditions.

It is worth noting that, under the АМ1.5 spectral conditions,
formula (12) turns into the expression
\begin{equation}
\eta_{\rm AM1.5}\approx 0.8 \; J_{\rm SC}(d) V_{\rm OC} FF/0.1,
\end{equation}
\noindent where we took a decrease of both the limit
short-circuit current and the irradiation  power under the АМ1.5
conditions into account.

\begin{figure}
\includegraphics[width=8.5cm]{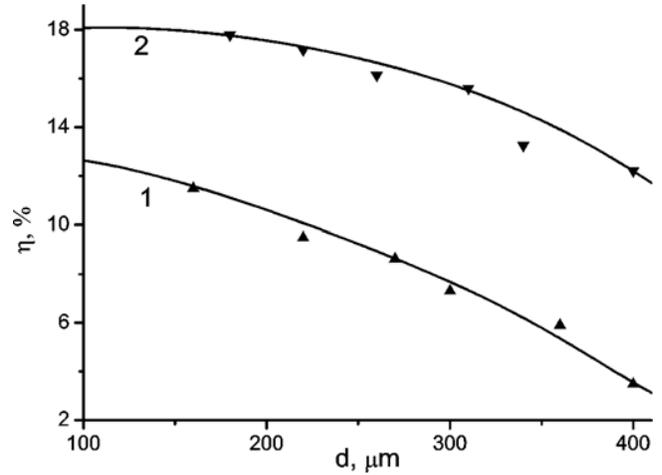}
\vskip-3mm\caption{Experimental (dots) and theoretical (solid lines)
thickness dependences of the BC SC photoconversion efficiency under
АМ1.5 conditions for the samples from the first ({\it 1}) and second
({\it 2}) groups. $R_d=0.7$, $L=500$ $\mu$m, $S_0=70$ cm/s, $d_0=70$
$\mu$m ({\it 1}), $R_d=0.7$, $L=250$ $\mu$m, $S_0=350$ cm/s,
$d_0=40$ $\mu$m ({\it 2})}
\end{figure}

Figure 6 presents the experimental thickness dependences of the
photoconversion  efficiency for the BC SC samples from the first and
second groups at room temperature under АМ1.5 conditions. The
theoretical dependences $\eta(d)$ were calculated using relation
(13), i.e, they were obtained with regard for the thickness dependence
$J_{\rm SC}(d)$ determined with the use of (10) and average values
of $V_{\rm OC}$ and $FF$. As one can see from Fig. 6, the
experimental thickness dependences of the photoconversion
efficiencies agree with the calculated ones, which justifies the use
of the thickness dependence $J_{\rm SC}(d)$ alone in (13). It is
worth noting that the agreement is reached using the diffusion
lengths for the samples from the first and second groups that were
determined with the help of expression (8) (see Fig.~3).

In addition, it turned out that the agreement between the
experimental and theoretical  functions $\eta(d)$ can be reached
only under the assumption that the effective surface recombination
rate depends on the thickness, namely, it was supposed that it
changes according to the law
\begin{equation}
S^*(d)=S_0\;\exp\left(\frac{d-d_{\max}}{d_0}\right),
\end{equation}
\noindent where $d_0$ is some parameter that determines a decrease of
$S^*$ with decrease in the thickness. In spite of small initial values of
$\eta$ for the samples from the first group  at $d=400$ $\mu$m,
which is related to the insufficiently large diffusion length, we
could realize the efficiency $\eta \approx 12 \%$ by reducing the
thickness of the studied sample to 160 $\mu$m and minimizing the
surface recombination rate due to the formation of a microporous
silicon layer. In the samples from the second group having the twice
larger diffusion length, the thinning results in the efficiency $\eta
\approx 18 \%$ (Fig.~6).

\section{Conclusions}

It is shown that, under certain conditions, the experimental
investigation of the  thickness dependences of the short-circuit
current of BC SCs in the case of strong light absorption and their
comparison with the theoretically calculated functions allow one to
determine both the surface recombination rate and the diffusion
length of minority carriers.

It is established that the experimental spectral dependences of the
short-circuit current of the studied BC SC samples agree with the
theoretical ones only with regard for the effect of light reflection
from the backside surface. If the value of $S^*$ is high, then the
position of their maximum practically does not depend on the
diffusion length $L$, whereas the value of $J_{\rm SC}$ in the
maximum significantly depends on $L$.

It is shown that, as the thickness decreases, the photoconversion
efficiency in the studied BC SC samples considerably grows.
Moreover, we have obtained a good agreement between the experimental and
theoretical dependences of $\eta(d)$.

\vskip-3mm
\rezume{%
ЗАЛЕЖНОСТІ ФОТОЕЛЕКТРИЧНИХ ХАРАКТЕРИСТИК КРЕМНІЄВИХ СОНЯЧНИХ
ЕЛЕМЕНТІВ З ТИЛОВОЮ МЕТАЛІЗАЦІЄЮ ВІД ТОВЩИНИ}{А.П. Горбань, В.П.
Костильов, А.В. Саченко, О.А. Серба,\\
  І.О.~Соколовський, В.В. Черненко} {Експериментально та теоретично досліджено товщинні залежності
квантового виходу фотоструму та фотоенергетичних параметрів
кремнієвих сонячних елементів з тиловою металізацією (СЕТМ).
Мінімізацію швидкості поверхневої рекомбінації (ШПР) на освітленій
поверхні в них досягнуто за рахунок створення шарів мікропористого
кремнію. Запропоновано метод знаходження  ШПР та довжини дифузії
неосновних носіїв заряду з товщинних залежностей квантового виходу
фотоструму в умовах сильного поглинання світла. Виконані дослідження
дозволили встановити, що потоншення зразків СЕТМ за умови
мінімізації ШПР дозволяє реалізувати достатньо великі значення
ефективності фотоперетворення. Показано також, що узгодження
експериментальних спектральних залежностей квантового виходу
фотоструму в досліджених СЕТМ з теоретичними може бути досягнуте
лише за умови врахування коефіцієнта відбиття світла від тилової
поверхні.}

\end{document}